\documentclass[a4paper,fleqn,usenatbib]{mnras}
\usepackage{graphicx}

\newcommand{\vsini}{\mbox{$v \sin i$}}

\newcommand{\sqiglt}{\hbox{\rlap{\lower.55ex \hbox {$\sim$}}
	\kern-.3em \raise.4ex \hbox{$<$}\,}}
\newcommand{\sqiggt}{\hbox{\rlap{\lower.55ex \hbox {$\sim$}}
	\kern-.3em \raise.4ex \hbox{$>$}\,}}

\title[WASP-166b]{WASP-166b: a bloated super-Neptune transiting a \mbox{$V$ = 9} star}

\author[Hellier et al.]{Coel Hellier$^{1}$, 
D.R. Anderson$^{1}$, 
A.H.M.J. Triaud$^{2,3}$,  
F. Bouchy$^{2}$,
A. Burdanov$^{4}$,\newauthor  
A. Collier Cameron$^{5}$,  
L. Delrez$^{4,6}$,
D. Ehrenreich$^{2}$,
M. Gillon$^{4}$,
E. Jehin$^{4}$,  \newauthor 
M. Lendl$^{7,2}$, 
E. Linder$^{8}$,    
L.D. Nielsen$^{2}$, 
P.F.L. Maxted$^{1}$,     
F. Pepe$^{2}$, 
D. Pollacco$^{9}$,  \newauthor   
D. Queloz$^{6}$,  
D. S\'egransan$^{2}$,   
B. Smalley$^{1}$, 
J. J. Spake$^{10}$, 
L. Y. Temple$^{1}$,
S. Udry$^{2}$,\newauthor 
R.G. West$^{9}$, and     
A. Wyttenbach$^{11}$\\
$^{1}$Astrophysics Group, Keele University, Staffordshire, ST5 5BG, UK\\
$^{2}$Observatoire astronomique de l'Universit\'e de Gen\`eve
51 ch. des Maillettes, 1290 Sauverny, Switzerland\\
$^{3}$School of Physics \&\ Astronomy, University of Birmingham, Edgbaston, Birmingham, B15 2TT, UK\\
$^{4}$Space sciences, Technologies and Astrophysics Research (STAR) Institute, Universit\'e de Li\`ege,\\  All\'ee du 6 Ao\^ut, 17, Bat. B5C, 4000 Li\`ege, Belgium\\
$^{5}$SUPA, School of Physics and Astronomy, University of St.\ Andrews, North Haugh,  Fife, KY16 9SS, UK\\
$^{6}$Cavendish Laboratory, J J Thomson Avenue, Cambridge, CB3 0HE, UK\\
$^{7}$Space Research Institute, Austrian Academy of Sciences, Schmiedlstr. 6, 8042, Graz, Austria\\
$^{8}$Physikalisches Institut, University of Bern, Sidlerstrasse 5, 3012, Bern, Switzerland\\
$^{9}$Department of Physics, University of Warwick, Gibbet Hill Road, Coventry CV4 7AL, UK\\
$^{10}$Astrophysics Group, School of Physics, University of Exeter, Stocker Road, Exeter, EX4 4QL, UK\\
$^{11}$Leiden Observatory, Leiden University, Postbus 9513, 2300 RA Leiden, The Netherlands}

\begin{document}

\date{date}
\pagerange{range}

\maketitle

\begin{abstract}
We report the discovery of WASP-166b, a super-Neptune planet with a mass of 0.1 M$_{\rm Jup}$ (1.9 M$_{\rm Nep}$) and a bloated radius of 0.63 R$_{\rm Jup}$.  It transits a $V$ = 9.36, F9V star in a 5.44-d orbit that is aligned with the stellar rotation axis (sky-projected obliquity angle $\lambda = 3 \pm 5$ degrees). Variations in the radial-velocity measurements are likely the result of magnetic activity over a 12-d stellar rotation period.  WASP-166b appears to be a rare object within the ``Neptune desert''. 
\end{abstract}


\section{Introduction}
Planets with low surface gravities have the largest atmospheric scale heights and so are the best targets for atmospheric characterisation by the technique of transmission spectroscopy, in which the planet's atmosphere is projected against the host-star photosphere during transit. Having a bloated radius also means that planets of sub-Saturn mass can still produce deep-enough transits to be found in ground-based surveys. Thus discoveries such as WASP-107b (0.12 M$_{\rm Jup}$; 0.94 R$_{\rm Jup}$; \citealt{2017A&A...604A.110A}) and WASP-127b (0.18 M$_{\rm Jup}$; 1.37 R$_{\rm Jup}$; \citealt{2017A&A...599A...3L}) are prime targets for  characterisation (e.g. \citealt{2018ApJ...858L...6K,2018Natur.557...68S,2017A&A...602L..15P,2018A&A...616A.145C}).  The importance of such targets, particularly ones transiting bright stars, will increase further with the launch of the {\it James Webb Space Telescope}. 

Planets between the masses of Neptune and Saturn are transitional between ice giants and gaseous giants, and so could help to elucidate why some proto-planets undergo runaway gaseous accretion while others do not. There are also far fewer Neptune-mass systems known, compared to the abundance of super-Earths found by the Kepler mission, and the several-hundred transiting hot Jupiters now found by the ground-based surveys.  The absence of Neptunes is particularly pronounced at short orbital periods, leading to the discussion of a ``Neptune desert'' \citep{2016A&A...589A..75M}. 

Here we report the discovery of WASP-166b, the lowest-mass planet yet found by the WASP survey, at only twice the mass of Neptune.

\begin{figure}
\hspace*{2mm}\includegraphics[width=8.5cm]{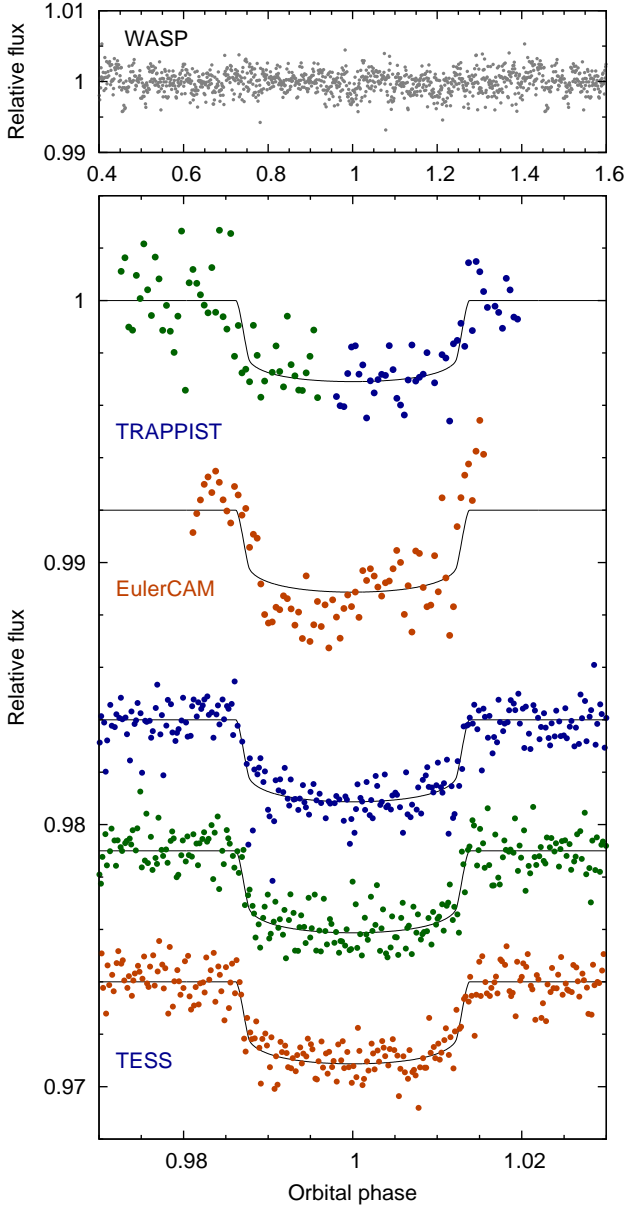}\\ [-2mm]
\caption{WASP-166b photometry: (Top) The WASP data folded on the 
transit period. (Main panel) Photometry from TRAPPIST-South, EulerCAM, and the three TESS transits, together with the fitted MCMC model (the TRAPPIST ingress, green points, and egress, blue points, are from different transits but are shown together).}
\end{figure}

\begin{figure}
\hspace*{2mm}\includegraphics[width=8.5cm]{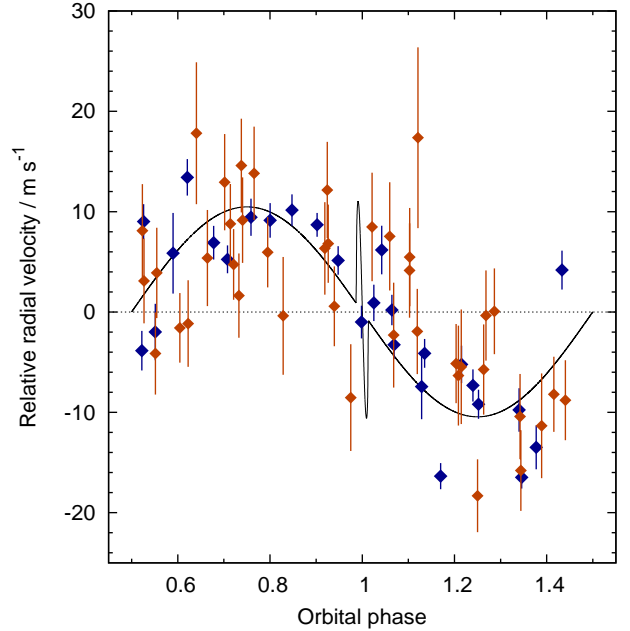}\\ [-2mm]
\caption{The HARPS (blue) and CORALIE (orange) radial velocities and fitted orbital model (the HARPS data in Fig.~4 are not shown here for clarity).}
\end{figure}

\begin{table}
\caption{Observations of WASP-166:\protect\rule[-1.0mm]{0mm}{1mm}}  
\begin{tabular}{lcr}
\hline 
Facility & Date & Notes \\ [0.5mm] \hline
WASP-South & 2006 May--2012 May & 33\,400 points \\ 
CORALIE  & 2014 Feb--2017 Jan &   41 RVs \\
HARPS (orbit) & 2016 Apr--2018 Mar &   27 RVs \\
HARPS (transit) & 2017 Jan 14 &   75 RVs \\
HARPS (transit) & 2017 Mar 04 &   52 RVs \\
HARPS (transit) & 2017 Mar 15 &   66 RVs \\
TRAPPIST-South & 2014 Mar 05 & $z$ band\\
EulerCAM  & 2016 Feb 05 & $I_{c}$ band \\ 
TRAPPIST-South & 2016 Feb  16 & $z^{\prime}$ band\\
TESS & 2019 Feb 2--27 & 3 transits 
\end{tabular} 
\end{table}

\section{Observations}
From 2006 to 2012 WASP-South operated as an array of  eight cameras based on 200-mm, f/1.8 Canon lenses backed by 2k$\times$2k Peltier-cooled CCDs (see  \citealt{2006PASP..118.1407P} for an account of the WASP project). Each field was observed with typically 10-min cadence.  The data were processed into a magnitude for each catalogued star, and the resulting lightcurves accumulated in a central archive. This was then searched for transit signals, with the best candidates sent for followup with the TRAPPIST-South photometer \citep{2013A&A...552A..82G} and the Euler/CORALIE spectrograph \citep{2013A&A...551A..80T}.  This combination has resulted in many discoveries of transiting exoplanets (e.g. \citealt{2018MNRAS.tmp.2617H}) and the techniques and methods used here are continuations of those from previous papers.

\begin{figure}
\hspace*{2mm}\includegraphics[width=8.5cm]{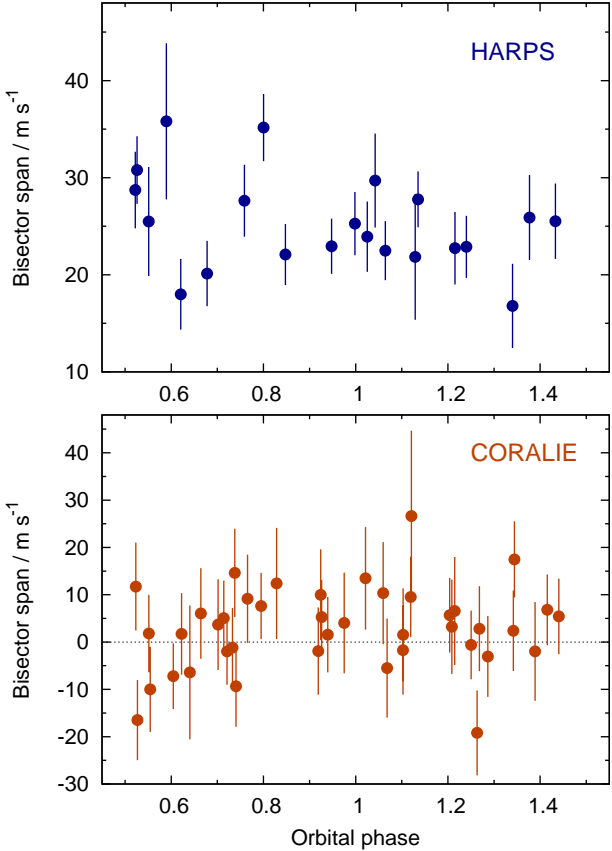}\\ [-2mm]
\caption{The spectroscopic bisector spans against orbital phase. The HARPS and CORALIE data are plotted separately since their accuracy is different.  The absence of any correlation with radial velocity is a check against transit mimics.}
\end{figure}

\begin{figure}
\hspace*{-5mm}\includegraphics[width=9cm]{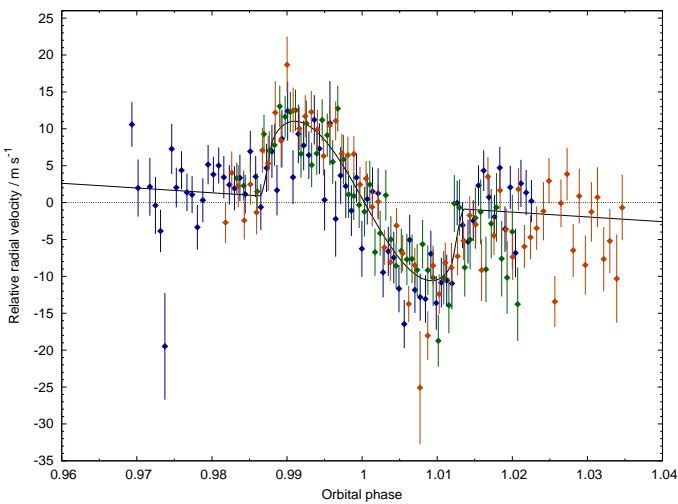}\\ [-2mm]
\caption{HARPS radial-velocity data through transit along with the fitted R--M model.  The colours denote data from different nights (blue: 2017-01-14; green 2017-03-04; orange 2017-03-15).} 
\end{figure}

\begin{figure}
\hspace*{-10mm}\includegraphics[width=10cm]{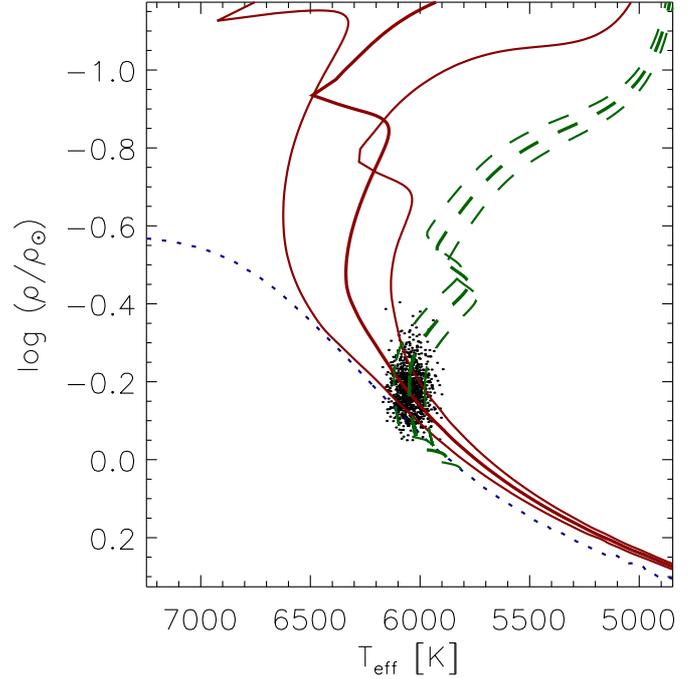}\\ [-2mm]
\caption{The host star's effective temperature (T$_{\rm eff}$) versus density (where the dots are outputs of the {\sc bagemass} MCMC). The blue dotted line is the zero-age main sequence, the green dashed lines are the evolutionary tracks (for the best-fitting mass of 1.18 M$_{\odot}$ and error bounds of 0.03 M$_{\odot}$), while the red lines are isochrones (for the best-fitting age of  2.1 Gyr and error bars of 0.9 Gyr).}
\label{trho_plot}
\end{figure}

WASP-166 was adopted as a candidate in 2014, after detection of a 5.44-d transit signal.   Radial-velocity (RV) observations with CORALIE found orbital motion at the transit period, but also showed additional variations.  We thus accumulated more RV data than is usual for WASP planet discoveries in order to look for additional bodies or a longer-term trend.   This amounted to 41 RVs with 1.2-m Euler/CORALIE over a three-year period, and 220 RVs with the ESO 3.6-m/HARPS, of which 27 covered the orbit, while 75, 52 and 66 were taken in sequences covering transits on three different nights.    The CORALIE and HARPS data were reduced using standard pipelines, as described in \citet{2019arXiv190401573R}, \citet{2019A&A...622A..37U}  and references therein. 

Further photometric observations (listed in Table~1) include two partial transit lightcurves from TRAPPIST-South (one ingress and one egress) and a full transit from EulerCAM \citep{2012A&A...544A..72L}, which unfortunately has excess red noise owing to poor observing conditions. 

In 2019 February, after completion of the initial version of this paper, the TESS satellite observed the sky sector that includes WASP-166.  TESS \citep{2016SPIE.9904E..2BR} is performing an all-sky transit survey aimed primarily at rocky planets too small to be found by the ground-based surveys. It has four cameras, each with a 10-cm lens backed by four 2048x2048 CCDs covering a field of 24$^{\circ}$x24$^{\circ}$, and observing in a bandpass from 600 to 1000 nm.  

We downloaded the public TESS lightcurve for WASP-166 (= TIC 408310006 = TOI-576) from the Mikulski Archive for Space Telescopes (MAST). We used the standard aperture-photometry data products, and extracted sections of the data around the known transit times. Three transits were observed over the 25-d period (a fourth was recorded shortly after recovery from a data gap, and we discard it owing to strong out-of-transit variability).

\section{Spectral analysis}
For a spectral analysis of the host star we combined and added the HARPS spectra and adopted the methods of \citet{2013MNRAS.428.3164D}. The resulting parameters are listed in Table~2. The effective temperature, $T_{\rm eff}$ = 6050 $\pm$ 50 K, comes from the H$\alpha$ line, and suggests a spectral type of F9. The surface gravity, $\log g$ = 4.5 $\pm$ 0.1 comes from Na~{\sc i} D and Mg~{\sc i} b lines.  The metallicity value of  [Fe/H] =  +0.19 $\pm$ 0.05 comes from  equivalent-width measurements of unblended Fe~{\sc i} lines.  The  Fe~{\sc i} lines were also used to estimate a rotation speed of $v \sin i$ = 4.6 $\pm$ 0.8 km\,s$^{-1}$, after convolving with the HARPS instrumental resolution ($R$ = 120\,000), and also accounting for  an estimate of the macroturbulence take from \citet{2014MNRAS.444.3592D}. We also report a value for the lithium abundance of log A(Li) =    2.68 $\pm$ 0.08.     

According to Gaia DR2 \citep{2018A&A...616A...1G}, WASP-166 has a relatively high proper motion of 56 mas/yr, which at the DR2 distance gives a transverse velocity of 30.0 $\pm$ 0.3 km\,s$^{-1}$, which complements the DR2 radial velocity of 24.0 $\pm$ 0.4 km\,s$^{-1}$. WASP-166  is relatively isolated, with the nearest DR2 star being 8 arcsecs away and 9 magnitudes fainter. There is no excess astrometric noise reported (such noise can indicate an unresolved binary). 

\begin{table}
\caption{System parameters for WASP-166.}  
\begin{tabular}{lc}
\multicolumn{2}{l}{1SWASP\,J093930.08--205856.8}\\
\multicolumn{2}{l}{2MASS\,09393009--2058568\ \ \ BD--20 2976}\\
\multicolumn{2}{l}{RA\,=\,09$^{\rm h}$39$^{\rm m}$30.09$^{\rm s}$, 
Dec\,=\,--20$^{\circ}$58$^{'}$56.9$^{''}$ (J2000)}\\
\multicolumn{2}{l}{$V$ mag = 9.36; GAIA $G$ = 9.26; $J$ = 8.35}  \\  
\multicolumn{2}{l}{Rotational modulation: $<$ 1 mmag}\\
\multicolumn{2}{l}{GAIA DR2 pm (RA) --55.082\,$\pm$\,0.072 (Dec) 10.927\,$\pm$\,0.069 mas/yr}\\
\multicolumn{2}{l}{GAIA DR2 parallax: 8.7301 $\pm$ 0.0448   mas}\\
\multicolumn{2}{l}{Distance = 113 $\pm$ 1 pc}\\
\hline
\multicolumn{2}{l}{Stellar parameters from spectroscopic analysis.\rule[-1.5mm]{0mm}{2mm}} \\ \hline 
Spectral type & F9V \\
$T_{\rm eff}$ (K)  & 6050  $\pm$ 50  \\
$\log g$      & 4.5 $\pm$ 0.1    \\
$v\,\sin i$ (km\,s$^{-1}$)     &   4.6 $\pm$ 0.8     \\
{[Fe/H]}   &  +0.19 $\pm$ 0.05     \\
log A(Li)  &    2.68 $\pm$ 0.08      \\ 
Age ({\sc bagemass}) (Gyr) & 2.1 $\pm$ 0.9 \\ \hline
\multicolumn{2}{l}{Parameters from MCMC analysis (fitted parameters denoted $\dagger$).\rule[-1.5mm]{0mm}{3mm}} \\
\hline 
$P$ (d)$^\dagger$ & 5.443540  $\pm$ 0.000004 \\
$T_{\rm c}$ (TDB)$^\dagger$ & 245\,7664.3289 $\pm$ 0.0006 \\
$T_{\rm 14}$ (d)$^\dagger$ & 0.150 $\pm$ 0.001 \\
$T_{\rm 12}=T_{\rm 34}$ (d) & 0.0088 $\pm$ 0.0012 \\
$\Delta F^\dagger = R_{\rm P}^{2}$/R$_{*}^{2}$ & 0.00281 $\pm$ 0.00007 \\
$b$$^\dagger$ & 0.39 $\pm$ 0.10 \\
$i$ ($^\circ$)  & 88.0 $\pm$ 0.7 \\
$K_{\rm 1}$ (km s$^{-1}$)$^\dagger$ & 0.0104 $\pm$ 0.0004 \\
$\gamma$ (km s$^{-1}$)$^\dagger$  & 23.6285 $\pm$ 0.0003 \\
$e$ & 0 (adopted) ($<$\,0.07 at 2$\sigma$) \\ 
$a/R_{\rm *}$  & 11.3 $\pm$ 0.6 \\ 
$M_{\rm *}$ (M$_{\rm \odot}$) & 1.19 $\pm$ 0.06 \\
$R_{\rm *}$ (R$_{\rm \odot}$) & 1.22 $\pm$ 0.06 \\
$\log g_{*}$ (cgs) & 4.34 $\pm$ 0.05 \\
$\rho_{\rm *}$ ($\rho_{\rm \odot}$) & 0.65 $\pm$ 0.10\\
$M_{\rm P}$ (M$_{\rm Jup}$) & 0.101 $\pm$ 0.005 \\
$R_{\rm P}$ (R$_{\rm Jup}$) & 0.63 $\pm$ 0.03 \\
$\log g_{\rm P}$ (cgs) & 2.77 $\pm$ 0.05 \\
$\rho_{\rm P}$ ($\rho_{\rm J}$) & 0.41 $\pm$ 0.07 \\
$\lambda$ (deg)$^\dagger$ & 3  $\pm$ 5 \\
$v \sin i$ (km s$^{-1}$) & 5.1 $\pm$ 0.3 \\
$a$ (AU)  & 0.0641  $\pm$ 0.0011 \\
Irradiation (W m$^{-2}$) & 6.0 $\pm$ 0.6 $\times 10^{5}$ \\ 
$T_{\rm P, A=0}$ (K) & 1270 $\pm$ 30 \\ [0.5mm] \hline 
\multicolumn{2}{l}{Priors were $M_{\rm *} = 1.18 \pm 0.03\ {\rm M}_{\odot}$ and
$R_{\rm *} = 1.23 \pm 0.06\ {\rm R}_{\odot}$}\\
\multicolumn{2}{l}{Errors are 1$\sigma$; Limb-darkening coefficients were:}\\
\multicolumn{2}{l}{{\small $r$ band: a1 = 0.512, a2 = 0.337, a3 = --0.138, 
a4 = --0.015}}\\ 
\multicolumn{2}{l}{{\small $I$ band: a1 = 0.579, a2 = 0.039, a3 = 0.104, 
a4 = --0.101}}\\ 
\multicolumn{2}{l}{{\small $z$ band: a1 =   0.599 , a2 = --0.076 , a3 = 0.191, a4 = --0.131}}\\ \hline
\end{tabular} 
\end{table}

\begin{figure}
\hspace*{-2mm}\includegraphics[width=9cm]{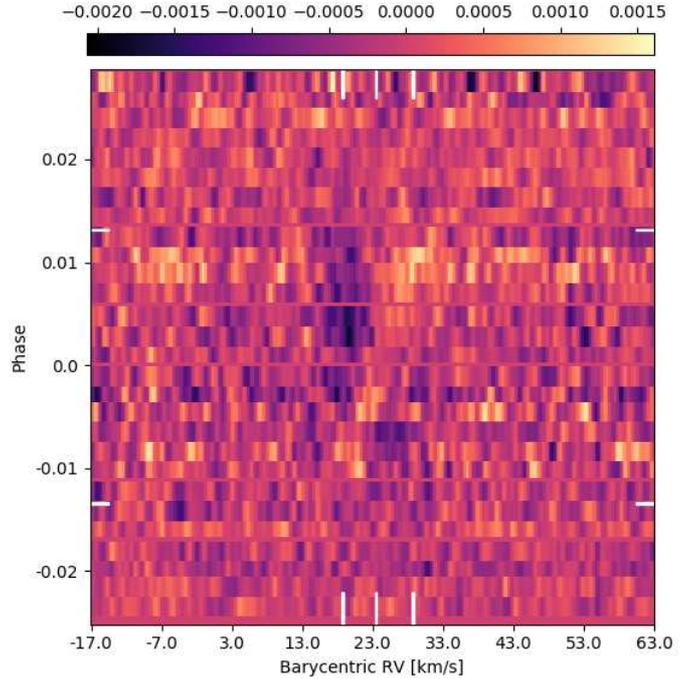}\\ [-2mm]
\caption{The line profiles through transit. The white lines show ($x$-axis) the mean $\gamma$ velocity of the system and the \vsini\ line width, and ($y$-axis) the beginning and end of transit.  The Doppler shadow of the planet moves from blue to red over the transit. The mean profile has been subtracted, resulting in a reduced level elsewhere in transit.}
\end{figure}

\section{System analysis}
As is standard for WASP discovery papers, we combined the photometry and radial-velocity datasets into a Markov-chain Monte-Carlo (MCMC) analysis (e.g.~\citealt{2007MNRAS.375..951C}). This fits parameters including $T_{\rm c}$ (the epoch of mid-transit),  $P$ (the orbital period), $\Delta F$ (the transit depth that would be observed in the absence of limb-darkening), $T_{14}$ (duration from first to fourth contact), $b$ (the impact parameter) and $K_{\rm 1}$ (the stellar reflex velocity).  For fitting the photometry we adopted the  4-parameter, non-linear limb darkening of \citet{2000A&A...363.1081C}, interpolating coefficients for the appropriate stellar temperature and metallicity. 

We allowed for radial-velocity offsets between different datasets, treating the CORALIE data before and after a November 2014 upgrade, the HARPS data round the orbit, and the three HARPS transit observations, all as independent sets (the RV values are listed in Table~A1).  The $\gamma$ velocity of the system given in Table~2 is that for the 27 HARPS datapoints around the orbit.  

We adopted a zero-eccentricity fit for WASP-166b, as is usually the case for lower-mass, short-period planets. It is clear, though, that there are additional radial-velocity deviations from the fitted model; indeed the fit to the HARPS RVs around the orbit has a $\chi^{2}$ of 217 for 27 datapoints.    Allowing an eccentric solution did not significantly improve the fit (producing a not-significant value of $e$ = 0.03 $\pm$ 0.02, with a 2-$\sigma$ upper limit of 0.07), nor did allowing a long-term drift in the RVs. Adding a second sinusoid, as could be caused by a second planet, also failed to model the additional variability, with tells us that it is not a coherent modulation.

The MCMC process accounts for the additional RV variability, and indeed any red noise in the transit lightcurves,  by inflating each dataset's errors to give  $\chi_{\nu}^{2} = 1$; this balances the different datasets' influence on the final result and inflates the errors on the output parameters.   

As is usual in WASP discovery papers we also constrained the stellar mass by adopting a prior based on the measured effective temperature and metallicity values, together with the stellar density obtained by an initial fit to the transit. For this we used the {\sc bagemass} code described in  \citet{2015A&A...575A..36M}. This resulted in a mass estimate of 1.18 $\pm$ 0.03 M$_{\odot}$, and also an age estimate of 2.1 $\pm$ 0.9 Gyr. The measured lithium abundance is consistent with this age estimate, though does not constrain the age further. 

Lastly, we include a constraint on the stellar radius derived from Gaia DR2 \citep{2018A&A...616A...1G}.   The DR2 parallax of  8.730 $\pm$ 0.045 mas implies a distance of 113 $\pm$ 1 pc (where we have applied the correction suggested by \citealt{2018ApJ...862...61S}).  Using the Infra-Red Flux Method \citep{1977MNRAS.180..177B} this implies a stellar radius of 1.23 $\pm$ 0.06 R$_{\odot}$, which we adopt as a prior.   The initial version of this paper lacked the TESS data and so, with only limited photometry, the parameters were less secure. However, the Gaia input tied down the stellar radius, and hence the impact parameter, and thus, after including the TESS data, the values have changed by less than an error bar. 

The adopted system parameters are listed in Table~2, while the data and fit are shown in Figs.~1 to 4. We also show a modified H--R diagram for the star in Fig.~\ref{trho_plot}.  

\begin{figure}
\hspace*{2mm}\includegraphics[width=8.5cm]{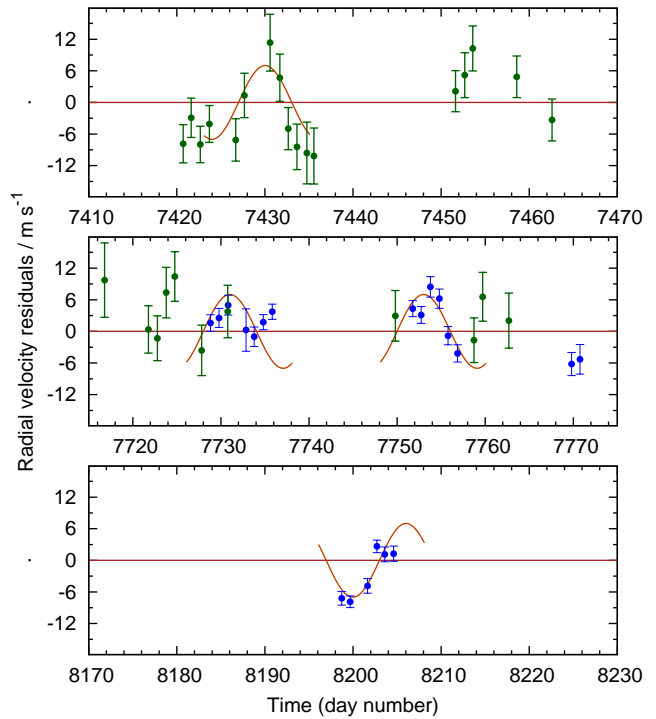}\\ [-2mm]
\caption{The residuals of the RV data to the orbital model, plotted as a function of time. Blue and green symbols are HARPS and CORALIE data respectively.  The red line shows portions of sinusoid (not phase coherent) that illustrate the putative 12.1-day stellar rotation period.}
\end{figure}

\subsection{The Rossiter--McLaughlin effect}
The above MCMC fit included fitting the in-transit radial velocities, where we use the parameterisation of \citet{2011ApJ...742...69H} to model the RV deviations caused by the Rossiter--McLaughlin effect (Fig.~4). Any differences between the data on the three different nights of data could be caused by the planet crossing starspots or faculae regions. 

The sky-projected obliquity angle, $\lambda$, is measured as \mbox{3 $\pm$ 5} degrees, and thus the planet's orbit is aligned with the stellar rotation.  The fitted \vsini\ of 5.1 $\pm$ 0.3 km\,s$^{-1}$ is consistent with the spectroscopic value of 4.6 $\pm$ 0.8 km\,s$^{-1}$. 

In Fig.~6 we show the HARPS line profiles through transit, in a tomographic display following the methods in \citet{2018MNRAS.480.5307T}. This figure shows the averaged data from all three transits, with the mean profile subtracted to better show variations. The planet trace can be seen moving prograde from blue to red over the transit.  Fitting the tomogram directly (e.g.~\citealt{2018MNRAS.480.5307T}), as opposed to fitting the modelled RVs, produces parameters consistent with those in Table~2, where again the uncertainties are currently dominated by the quality of the photometry.

\subsection{Possible magnetic activity}
The RV data clearly show deviations about the orbital model (Fig.~2). To illustrate this we plot in Fig.~7 some of the residual RV values as a function of time  (omitting the in-transit R--M sequences).   The residuals appear to be correlated from night to night. 

Given the transit and the fact that the planet's orbit is aligned we can presume that the stellar rotation axis is perpendicular to the line of sight, and thus combining the \vsini\ fitted to the R--M effect with the fitted stellar radius we obtain a rotational period of 12.1 $\pm$ 0.9 days. To guide the eye we plot in Fig.~7  portions of 12.1-d sinusoid (amplitude 7 m s$^{-1}$), and conclude that the RV deviations might result from magnetic activity. 

We have also investigated the variability by modelling the HARPS RVs using a gaussian-process (GP) analysis following the method detailed in \citet{2014MNRAS.443.2517H}.  To model the residuals, this adds to the orbital motion three hyperparameters (with uniform priors), namely an amplitude $a$, a period $P_{\rm rot}$ (taken to be the rotational period), and an exponential decay timescale, $t_{\rm decay}$ (being the timescale on which magnetic activity would lose coherence). We set the last to $\sim$\,30 d, which is physically realistic, but is not constrained by the data since we have few observations separated by that timescale.   The ``corner plot'' of the GP output is shown in Fig.~8.

The analysis produces a clear preference for a rotational period of 12.3 $\pm$ 1.9 days, consistent with that from the R--M  \vsini, and in line with the magnetic-activity hypothesis.  The resulting values of $K_{1}$ = 0.0100 $\pm$ 0.0006 km\,s$^{-1}$ and $\gamma$ = 23.6288 $\pm$ 0.0012 km\,s$^{-1}$ are consistent with those in Table~2. 

\begin{figure}
\vspace*{2mm}\hspace*{-5mm}\includegraphics[width=9cm]{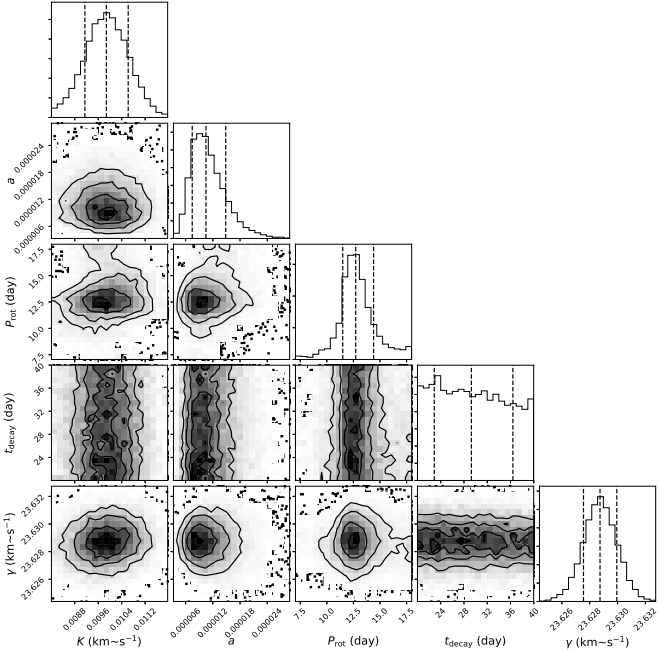}\\ [0mm]
\caption{Parameter distributions from the gaussian-process analysis of the correlated RV residuals. The decay timescale $t_{\rm decay}$ is not constrained by the data, but there is a clear preference for a rotational period, $P_{\rm rot}$ = 12.3 $\pm$ 1.9 day.}
\end{figure}

Having found evidence of a 12-d rotational period we then searched the WASP lightcurve for any rotational modulation, using the methods of \citet{2011PASP..123..547M}. The WASP data amount to 33\,000 data points spanning $\sim$\,150 nights of coverage in each of six consecutive seasons from 2007 to 2012. We found no significant modulation to a 95\%\ limit of 1 mmag.   Given the lack of a photometric modulation, the correlated RV residuals might be attributable to magnetic suppression of photospheric convection in spot-free facular active regions (e.g.~\citealt{2019ApJ...874..107M}).

\begin{figure}
\hspace*{0mm}\includegraphics[width=9cm]{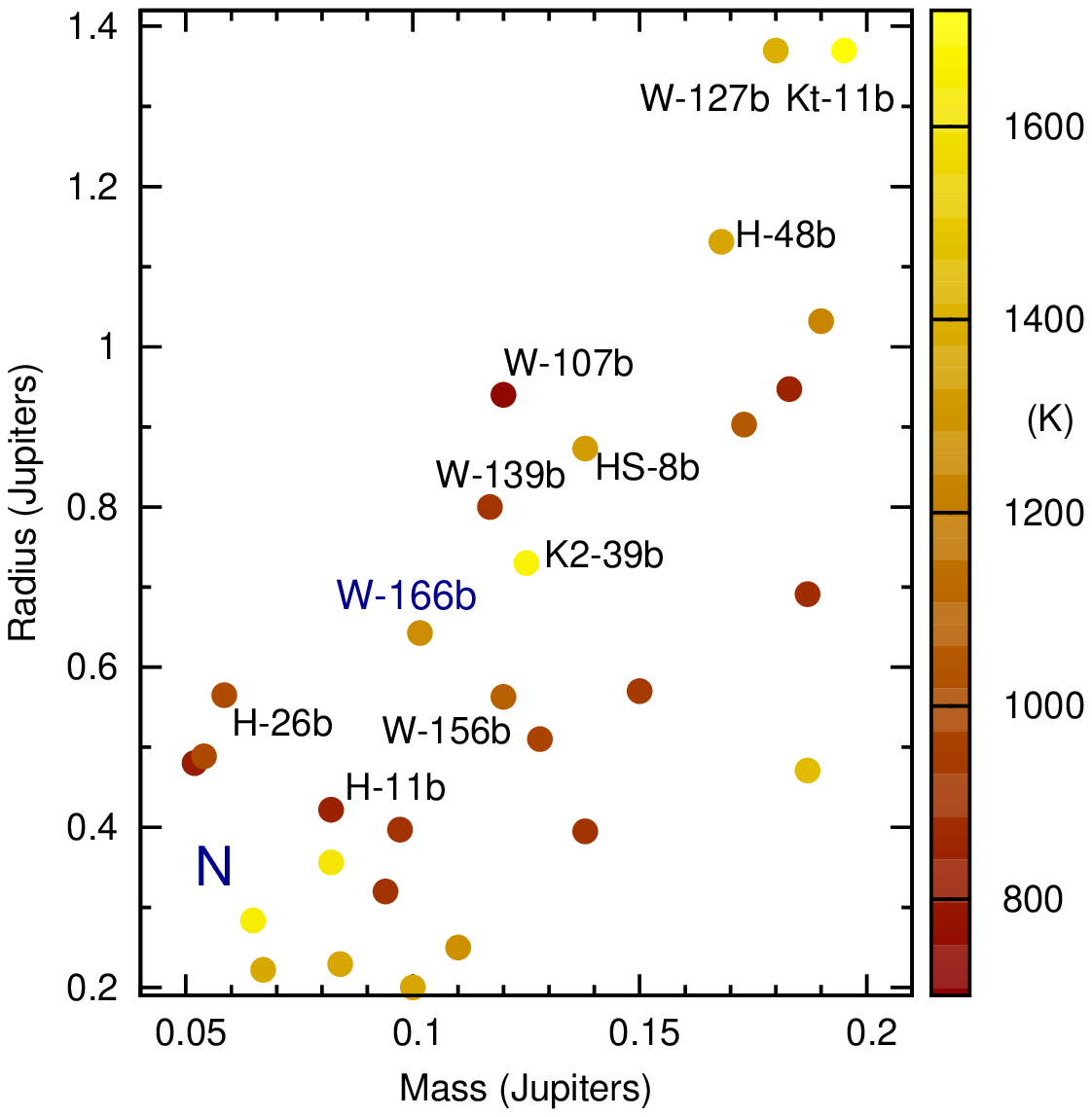}\\ [-2mm]
\caption{Masses and radii of transiting ``hot'' super-Neptune planets (with orbital periods $<$\,10 d). The symbols are coloured according to the planet's equilibrium temperature.  The labelled planets are WASP-107b \citep{2017A&A...604A.110A}, WASP-127b \citep{2017A&A...599A...3L}, WASP-139b \citep{2017MNRAS.465.3693H}, WASP-156b \citep{2018A&A...610A..63D}, HAT-P-11b \citep{2010ApJ...710.1724B}, HAT-P-26b \citep{2011ApJ...728..138H}, HAT-P-48b \citep{2016arXiv160604556B}, HATS-8b \citep{2015AJ....150...49B}, KELT-11b \citep{2017AJ....153..215P} and K2-39b \citep{2016AJ....152..143V,2017AJ....153..142P}. The location of Neptune is marked with an N.}
\end{figure}

\section{Discussion}
With a mass of 1.9 Neptunes, WASP-166b is the lowest-mass planet yet discovered by the WASP survey. It also has a bloated radius of 0.63 $\pm$ 0.03 R$_{\rm Jup}$.   The ``super-Neptune'' region of the mass--radius diagram for known, short-period transiting exoplanets is plotted in Fig.~9, showing that WASP-166b is near the upper size bound for a planet of its mass. The empirically observed upper bound (from WASP-127b to WASP-107b to HAT-P-26b) is falling rapidly in this mass range, which may be telling us about radius-inflation mechanisms and the ability of a lower-mass planet to hold onto its envelope under the effects of irradiation.     

For example, \citet{2016A&A...589A..75M} have shown that there is a ``Neptune desert'' at short orbital periods, with almost no hot-Neptune planets at periods $<$ 5 d and fewer at periods $<$ 10 d, when compared to abundant hot Jupiters and super-Earths (see also \citealt{2018AJ....155..214T} on the lack of inflated sub-Saturns). To illustrate the ``Neptune'' or ``sub-Jovian desert'' we highlight the location of WASP-166b on a plot of planet mass against insolation (Fig.~10). 

The desert is likely the result of photo-irradiation of inwardly migrating planets (e.g.~\citealt{2018A&A...616A..76S,2018MNRAS.479.5012O,2019arXiv190304817S}). Jupiter-mass planets are able to resist photo-evaporation, and continue to migrate inwards by tidal orbital  decay, whereas a low-surface-gravity Neptune such as WASP-166b could not.  A hot Neptune could instead be captured into a short-period orbit from a high-eccentricity-migration pathway, and so, being new to its orbit, would not have undergone the photo-evaporation that would have occurred had it migrated there by the slower process of orbital decay. However, such capture only occurs for a narrow range of parameter space \citep{2018MNRAS.479.5012O}, and so such systems would be rare.  

\begin{figure*}
\vspace*{2mm}\hspace*{-10mm}\includegraphics[width=16cm]{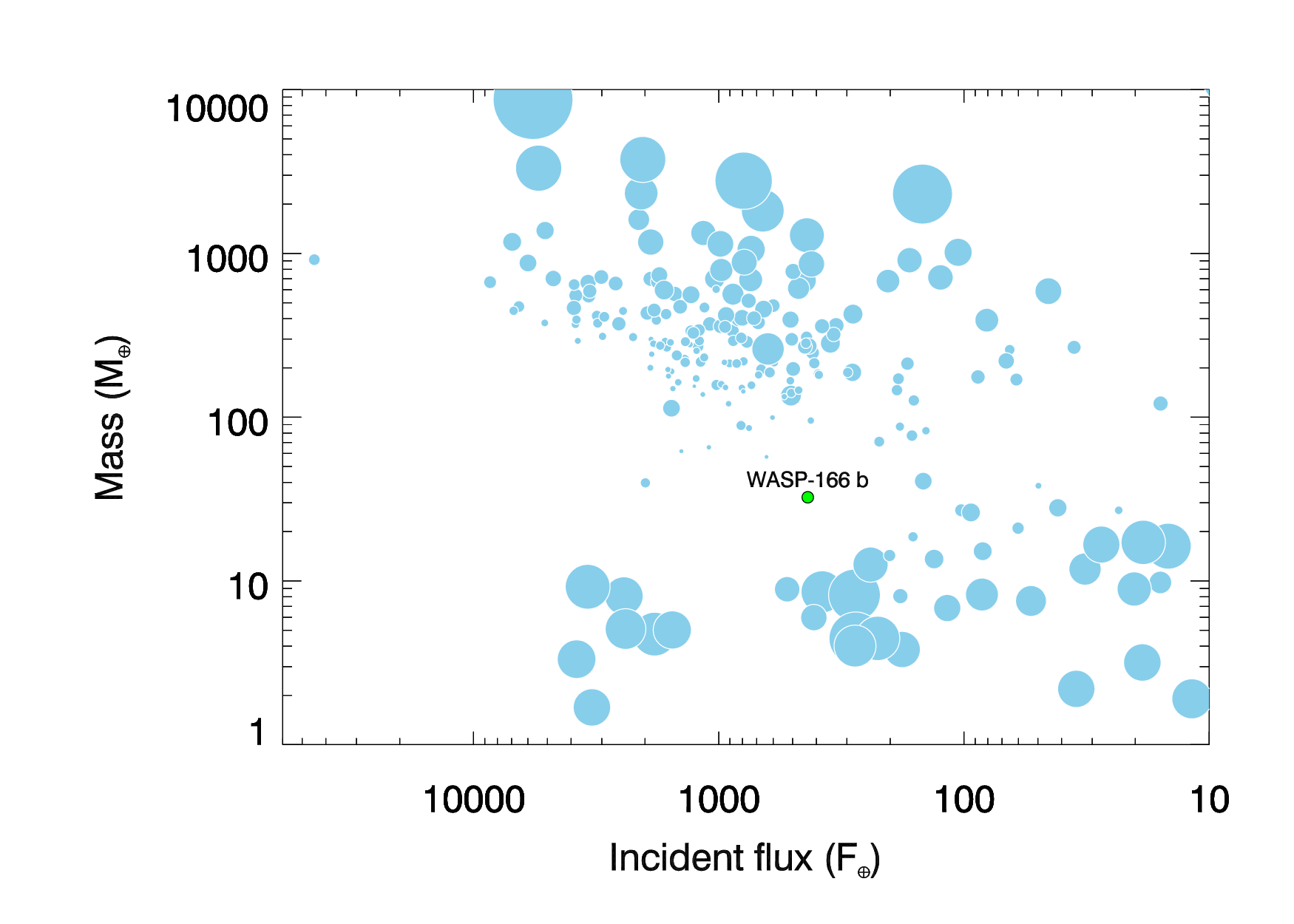}\\ [0mm]
\caption{Exoplanet mass versus incident flux, showing the location of WASP-166b in the ``sub-Jovian desert''. The symbol area is proportional to the bulk density of the planet. We include only planets with host stars brighter than $V = 12$. Data are from http://exoplanets.org/}
\end{figure*}

With a period of 5.4 d and orbiting an F star, WASP-166b has a relatively high irradiation of 6 $\times$ 10$^{5}$ W m$^{-2}$ (440 times Earth's insolation) for such a low-surface-gravity planet.  Thus it appears to be a rare object, with a radius bloated by irradiation, on the boundary of the Neptune desert.

The Rossiter--McLaughlin observations show that the orbit is aligned ($\lambda$ = 3  $\pm$ 5 degrees). This is consistent with it being on a lengthy inwardly migration, during which it has become bloated owing to irradiation. However, recent capture from high-eccentricity pathways can also produce aligned orbits, so this is not conclusive.  Relatively few of the super-Neptune planets in Fig.~9 have had obliquity angles measured, though those that have -- HAT-P-11b \citep{2010ApJ...710.1724B,2010ApJ...723L.223W}, WASP-107b \citep{2017A&A...604A.110A,2017AJ....153..205D,2017MNRAS.469.1622M} and GJ\,436b \citep{2007A&A...472L..13G,2018Natur.553..477B} -- are all known to be misaligned. 

The bloated nature of WASP-166b combines with a bright host star of $V$ = 9.36 to make it a prime target for atmospheric characterisation. Indeed, recent observations of irradiated low-surface-gravity planets show indications of photo-evaporating atmospheres (e.g.~\citealt{2018Natur.557...68S,2018ApJ...868L..34M}).  The expected signal for a transmission spectrum depends on the atmospheric scale height, transit depth, and host-star magnitude (e.g. equation 36 of \citealt{2010arXiv1001.2010W}). In Fig.~11 we compare the signal expected for WASP-166b with other low-mass planets ($M < 0.2$ M$_{\rm Jup}$). This shows that WASP-166b is likely to be among the best targets for such studies, though this may be more difficult if the star is indeed magnetically active, as indicated by correlated deviations in the RV data.  We suggest that WASP-166b is potentially a prime target for the {\it James Webb Space Telescope}, and that it should first be observed with {\it HST\/} in order to assess the cloudiness of its atmosphere.

\begin{figure*}
\vspace*{2mm}\hspace*{10mm}\includegraphics[width=10cm,angle=-90]{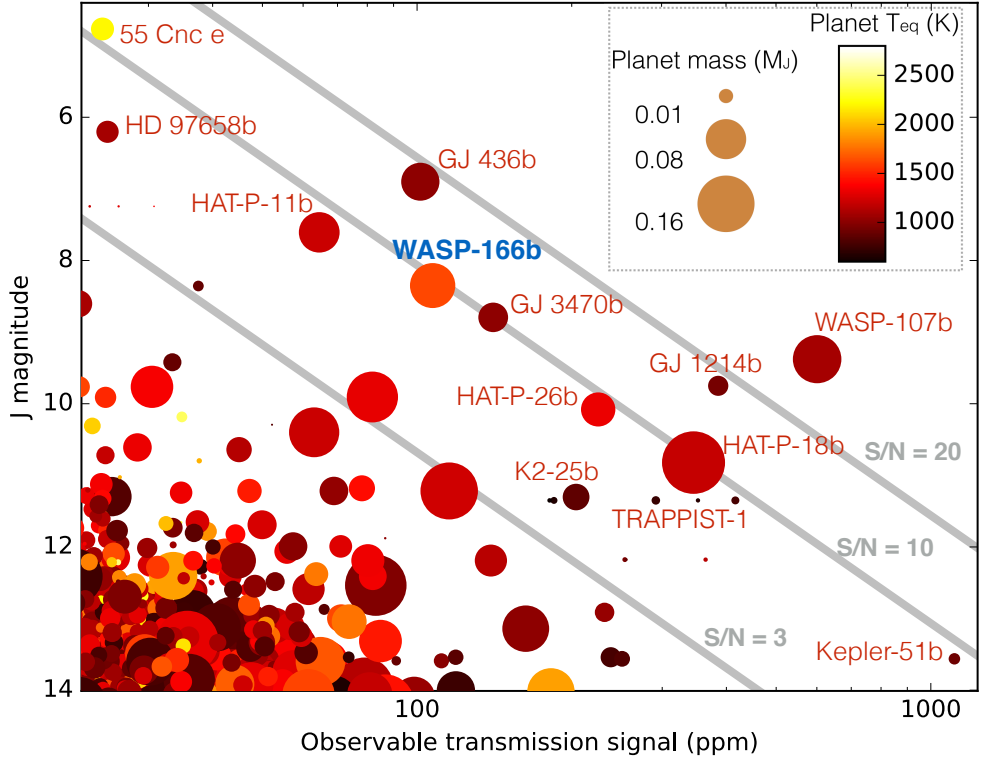}\\ [0mm]
\caption{An illustration of prime low-mass planets ($< 0.2$ M$_{\rm Jup}$) for atmospheric characterisation, based on the scale height of the atmosphere, the transit depth, and the host-star brightness.}
\end{figure*}

\section*{Acknowledgements}
WASP-South is hosted by the South African Astronomical Observatory and
we are grateful for their ongoing support and assistance. Funding for
WASP came from consortium universities and from the UK's Science and
Technology Facilities Council (STFC). ACC acknowledges support from STFC consolidated grant number ST/R000824/1.  The Euler Swiss telescope is supported
by the Swiss National Science Foundation. The HARPS data result from observations made at ESO 3.6~m telescope at the La Silla Observatory under ESO programmes 097.C-0434 and 098.C-0304.  TRAPPIST-South is funded by
the Belgian Fund for Scientific Research (Fond National de la
Recherche Scientifique, FNRS) under the grant FRFC 2.5.594.09.F, with
the participation of the Swiss National Science Fundation (SNF). The
research leading to these results has received funding from the ARC
grant for Concerted Research Actions, financed by the
Wallonia-Brussels Federation. M.G. and E.J. are Senior Research Associates at the FNRS-F.R.S. L.D. acknowledges support from a Gruber Foundation Fellowship.  This project has received funding from the European Research Council (ERC) under the European Union’s Horizon 2020 research and innovation programme (project {\sc Four Aces}; grant agreement No 724427). This work has been carried out in the frame of the National Centre for Competence in Research PlanetS supported by the Swiss National Science Foundation (SNSF). This paper includes data collected with the TESS mission, obtained from the MAST data archive at the Space Telescope Science Institute (STScI). Funding for the TESS mission is provided by the NASA Explorer Program.  We thank the many people involved in the creation of TESS data.  We acknowledge use of data from the European Space Agency (ESA) mission Gaia, as processed by the Gaia Data Processing and Analysis Consortium (DPAC). 

\bibliographystyle{mnras}
\bibliography{biblio}

\begin{thebibliography}{}
\makeatletter
\relax
\def\mn@urlcharsother{\let\do\@makeother \do\$\do\&\do\#\do\^\do\_\do\%\do\~}
\def\mn@doi{\begingroup\mn@urlcharsother \@ifnextchar [ {\mn@doi@}
  {\mn@doi@[]}}
\def\mn@doi@[#1]#2{\def\@tempa{#1}\ifx\@tempa\@empty \href
  {http://dx.doi.org/#2} {doi:#2}\else \href {http://dx.doi.org/#2} {#1}\fi
  \endgroup}
\def\mn@eprint#1#2{\mn@eprint@#1:#2::\@nil}
\def\mn@eprint@arXiv#1{\href {http://arxiv.org/abs/#1} {{\tt arXiv:#1}}}
\def\mn@eprint@dblp#1{\href {http://dblp.uni-trier.de/rec/bibtex/#1.xml}
  {dblp:#1}}
\def\mn@eprint@#1:#2:#3:#4\@nil{\def\@tempa {#1}\def\@tempb {#2}\def\@tempc
  {#3}\ifx \@tempc \@empty \let \@tempc \@tempb \let \@tempb \@tempa \fi \ifx
  \@tempb \@empty \def\@tempb {arXiv}\fi \@ifundefined
  {mn@eprint@\@tempb}{\@tempb:\@tempc}{\expandafter \expandafter \csname
  mn@eprint@\@tempb\endcsname \expandafter{\@tempc}}}

\bibitem[\protect\citeauthoryear{{Anderson} et~al.,}{{Anderson}
  et~al.}{2017}]{2017A&A...604A.110A}
{Anderson} D.~R.,  et~al., 2017, \mn@doi [\aap] {10.1051/0004-6361/201730439},
  \href {http://adsabs.harvard.edu/abs/2017A%26A...604A.110A} {604, A110}

\bibitem[\protect\citeauthoryear{{Bakos} et~al.,}{{Bakos}
  et~al.}{2010}]{2010ApJ...710.1724B}
{Bakos} G.~{\'A}.,  et~al., 2010, \mn@doi [\apj]
  {10.1088/0004-637X/710/2/1724}, \href
  {http://adsabs.harvard.edu/abs/2010ApJ...710.1724B} {710, 1724}

\bibitem[\protect\citeauthoryear{{Bakos} et~al.,}{{Bakos}
  et~al.}{2016}]{2016arXiv160604556B}
{Bakos} G.~{\'A}.,  et~al., 2016, preprint, \href
  {http://adsabs.harvard.edu/abs/2016arXiv160604556B} {} (\mn@eprint {arXiv}
  {1606.04556})

\bibitem[\protect\citeauthoryear{{Bayliss} et~al.,}{{Bayliss}
  et~al.}{2015}]{2015AJ....150...49B}
{Bayliss} D.,  et~al., 2015, \mn@doi [\aj] {10.1088/0004-6256/150/2/49}, \href
  {http://adsabs.harvard.edu/abs/2015AJ....150...49B} {150, 49}

\bibitem[\protect\citeauthoryear{{Blackwell} \& {Shallis}}{{Blackwell} \&
  {Shallis}}{1977}]{1977MNRAS.180..177B}
{Blackwell} D.~E.,  {Shallis} M.~J.,  1977, \mn@doi [\mnras]
  {10.1093/mnras/180.2.177}, \href
  {http://adsabs.harvard.edu/abs/1977MNRAS.180..177B} {180, 177}

\bibitem[\protect\citeauthoryear{{Bourrier} et~al.,}{{Bourrier}
  et~al.}{2018}]{2018Natur.553..477B}
{Bourrier} V.,  et~al., 2018, \mn@doi [\nat] {10.1038/nature24677}, \href
  {http://adsabs.harvard.edu/abs/2018Natur.553..477B} {553, 477}

\bibitem[\protect\citeauthoryear{{Chen} et~al.,}{{Chen}
  et~al.}{2018}]{2018A&A...616A.145C}
{Chen} G.,  et~al., 2018, \mn@doi [\aap] {10.1051/0004-6361/201833033}, \href
  {http://adsabs.harvard.edu/abs/2018A%26A...616A.145C} {616, A145}

\bibitem[\protect\citeauthoryear{{Claret}}{{Claret}}{2000}]{2000A&A...363.1081C}
{Claret} A.,  2000, \aap, \href
  {http://adsabs.harvard.edu/abs/2000A%26A...363.1081C} {363, 1081}

\bibitem[\protect\citeauthoryear{{Collier Cameron} et~al.,}{{Collier Cameron}
  et~al.}{2007}]{2007MNRAS.375..951C}
{Collier Cameron} A.,  et~al., 2007, \mn@doi [\mnras]
  {10.1111/j.1365-2966.2006.11350.x}, \href
  {http://adsabs.harvard.edu/abs/2007MNRAS.375..951C} {375, 951}

\bibitem[\protect\citeauthoryear{{Dai} \& {Winn}}{{Dai} \&
  {Winn}}{2017}]{2017AJ....153..205D}
{Dai} F.,  {Winn} J.~N.,  2017, \mn@doi [\aj] {10.3847/1538-3881/aa65d1}, \href
  {http://adsabs.harvard.edu/abs/2017AJ....153..205D} {153, 205}

\bibitem[\protect\citeauthoryear{{Demangeon} et~al.,}{{Demangeon}
  et~al.}{2018}]{2018A&A...610A..63D}
{Demangeon} O.~D.~S.,  et~al., 2018, \mn@doi [\aap]
  {10.1051/0004-6361/201731735}, \href
  {http://adsabs.harvard.edu/abs/2018A%26A...610A..63D} {610, A63}

\bibitem[\protect\citeauthoryear{{Doyle} et~al.,}{{Doyle}
  et~al.}{2013}]{2013MNRAS.428.3164D}
{Doyle} A.~P.,  et~al., 2013, \mn@doi [\mnras] {10.1093/mnras/sts267}, \href
  {http://adsabs.harvard.edu/abs/2013MNRAS.428.3164D} {428, 3164}

\bibitem[\protect\citeauthoryear{{Doyle}, {Davies}, {Smalley}, {Chaplin}  \&
  {Elsworth}}{{Doyle} et~al.}{2014}]{2014MNRAS.444.3592D}
{Doyle} A.~P.,  {Davies} G.~R.,  {Smalley} B.,  {Chaplin} W.~J.,   {Elsworth}
  Y.,  2014, \mn@doi [\mnras] {10.1093/mnras/stu1692}, \href
  {http://adsabs.harvard.edu/abs/2014MNRAS.444.3592D} {444, 3592}

\bibitem[\protect\citeauthoryear{{Gaia Collaboration} et~al.,}{{Gaia
  Collaboration} et~al.}{2018}]{2018A&A...616A...1G}
{Gaia Collaboration} et~al., 2018, \mn@doi [\aap]
  {10.1051/0004-6361/201833051}, \href
  {http://adsabs.harvard.edu/abs/2018A%26A...616A...1G} {616, A1}

\bibitem[\protect\citeauthoryear{{Gillon} et~al.,}{{Gillon}
  et~al.}{2007}]{2007A&A...472L..13G}
{Gillon} M.,  et~al., 2007, \mn@doi [\aap] {10.1051/0004-6361:20077799}, \href
  {http://adsabs.harvard.edu/abs/2007A%26A...472L..13G} {472, L13}

\bibitem[\protect\citeauthoryear{{Gillon} et~al.,}{{Gillon}
  et~al.}{2013}]{2013A&A...552A..82G}
{Gillon} M.,  et~al., 2013, \mn@doi [\aap] {10.1051/0004-6361/201220561}, \href
  {http://adsabs.harvard.edu/abs/2013A%26A...552A..82G} {552, A82}

\bibitem[\protect\citeauthoryear{{Hartman} et~al.,}{{Hartman}
  et~al.}{2011}]{2011ApJ...728..138H}
{Hartman} J.~D.,  et~al., 2011, \mn@doi [\apj] {10.1088/0004-637X/728/2/138},
  \href {http://adsabs.harvard.edu/abs/2011ApJ...728..138H} {728, 138}

\bibitem[\protect\citeauthoryear{{Haywood} et~al.,}{{Haywood}
  et~al.}{2014}]{2014MNRAS.443.2517H}
{Haywood} R.~D.,  et~al., 2014, \mn@doi [\mnras] {10.1093/mnras/stu1320}, \href
  {http://adsabs.harvard.edu/abs/2014MNRAS.443.2517H} {443, 2517}

\bibitem[\protect\citeauthoryear{{Hellier} et~al.,}{{Hellier}
  et~al.}{2017}]{2017MNRAS.465.3693H}
{Hellier} C.,  et~al., 2017, \mn@doi [\mnras] {10.1093/mnras/stw3005}, \href
  {http://adsabs.harvard.edu/abs/2017MNRAS.465.3693H} {465, 3693}

\bibitem[\protect\citeauthoryear{{Hellier} et~al.,}{{Hellier}
  et~al.}{2018}]{2018MNRAS.tmp.2617H}
{Hellier} C.,  et~al., 2018, \mn@doi [\mnras] {10.1093/mnras/sty2741}, \href
  {http://adsabs.harvard.edu/abs/2018MNRAS.tmp.2617H} {}

\bibitem[\protect\citeauthoryear{{Hirano}, {Suto}, {Winn}, {Taruya}, {Narita},
  {Albrecht}  \& {Sato}}{{Hirano} et~al.}{2011}]{2011ApJ...742...69H}
{Hirano} T.,  {Suto} Y.,  {Winn} J.~N.,  {Taruya} A.,  {Narita} N.,  {Albrecht}
  S.,   {Sato} B.,  2011, \mn@doi [\apj] {10.1088/0004-637X/742/2/69}, \href
  {http://adsabs.harvard.edu/abs/2011ApJ...742...69H} {742, 69}

\bibitem[\protect\citeauthoryear{{Kreidberg}, {Line}, {Thorngren}, {Morley}  \&
  {Stevenson}}{{Kreidberg} et~al.}{2018}]{2018ApJ...858L...6K}
{Kreidberg} L.,  {Line} M.~R.,  {Thorngren} D.,  {Morley} C.~V.,   {Stevenson}
  K.~B.,  2018, \mn@doi [\apjl] {10.3847/2041-8213/aabfce}, \href
  {http://adsabs.harvard.edu/abs/2018ApJ...858L...6K} {858, L6}

\bibitem[\protect\citeauthoryear{{Lam} et~al.,}{{Lam}
  et~al.}{2017}]{2017A&A...599A...3L}
{Lam} K.~W.~F.,  et~al., 2017, \mn@doi [\aap] {10.1051/0004-6361/201629403},
  \href {http://adsabs.harvard.edu/abs/2017A%26A...599A...3L} {599, A3}

\bibitem[\protect\citeauthoryear{{Lendl} et~al.,}{{Lendl}
  et~al.}{2012}]{2012A&A...544A..72L}
{Lendl} M.,  et~al., 2012, \mn@doi [\aap] {10.1051/0004-6361/201219585}, \href
  {http://adsabs.harvard.edu/abs/2012A%26A...544A..72L} {544, A72}

\bibitem[\protect\citeauthoryear{{Mansfield} et~al.,}{{Mansfield}
  et~al.}{2018}]{2018ApJ...868L..34M}
{Mansfield} M.,  et~al., 2018, \mn@doi [\apjl] {10.3847/2041-8213/aaf166},
  \href {http://adsabs.harvard.edu/abs/2018ApJ...868L..34M} {868, L34}

\bibitem[\protect\citeauthoryear{{Maxted} et~al.,}{{Maxted}
  et~al.}{2011}]{2011PASP..123..547M}
{Maxted} P.~F.~L.,  et~al., 2011, \mn@doi [\pasp] {10.1086/660007}, \href
  {http://adsabs.harvard.edu/abs/2011PASP..123..547M} {123, 547}

\bibitem[\protect\citeauthoryear{{Maxted}, {Serenelli}  \&
  {Southworth}}{{Maxted} et~al.}{2015}]{2015A&A...575A..36M}
{Maxted} P.~F.~L.,  {Serenelli} A.~M.,   {Southworth} J.,  2015, \mn@doi [\aap]
  {10.1051/0004-6361/201425331}, \href
  {http://adsabs.harvard.edu/abs/2015A%26A...575A..36M} {575, A36}

\bibitem[\protect\citeauthoryear{{Mazeh}, {Holczer}  \& {Faigler}}{{Mazeh}
  et~al.}{2016}]{2016A&A...589A..75M}
{Mazeh} T.,  {Holczer} T.,   {Faigler} S.,  2016, \mn@doi [\aap]
  {10.1051/0004-6361/201528065}, \href
  {http://adsabs.harvard.edu/abs/2016A%26A...589A..75M} {589, A75}

\bibitem[\protect\citeauthoryear{{Milbourne} et~al.,}{{Milbourne}
  et~al.}{2019}]{2019ApJ...874..107M}
{Milbourne} T.~W.,  et~al., 2019, \mn@doi [\apj] {10.3847/1538-4357/ab064a},
  \href {http://adsabs.harvard.edu/abs/2019ApJ...874..107M} {874, 107}

\bibitem[\protect\citeauthoryear{{Mo{\v c}nik}, {Hellier}, {Anderson}, {Clark}
  \& {Southworth}}{{Mo{\v c}nik} et~al.}{2017}]{2017MNRAS.469.1622M}
{Mo{\v c}nik} T.,  {Hellier} C.,  {Anderson} D.~R.,  {Clark} B.~J.~M.,
  {Southworth} J.,  2017, \mn@doi [\mnras] {10.1093/mnras/stx972}, \href
  {http://adsabs.harvard.edu/abs/2017MNRAS.469.1622M} {469, 1622}

\bibitem[\protect\citeauthoryear{{Owen} \& {Lai}}{{Owen} \&
  {Lai}}{2018}]{2018MNRAS.479.5012O}
{Owen} J.~E.,  {Lai} D.,  2018, \mn@doi [\mnras] {10.1093/mnras/sty1760}, \href
  {http://adsabs.harvard.edu/abs/2018MNRAS.479.5012O} {479, 5012}

\bibitem[\protect\citeauthoryear{{Palle} et~al.,}{{Palle}
  et~al.}{2017}]{2017A&A...602L..15P}
{Palle} E.,  et~al., 2017, \mn@doi [\aap] {10.1051/0004-6361/201731018}, \href
  {http://adsabs.harvard.edu/abs/2017A%26A...602L..15P} {602, L15}

\bibitem[\protect\citeauthoryear{{Pepper} et~al.,}{{Pepper}
  et~al.}{2017}]{2017AJ....153..215P}
{Pepper} J.,  et~al., 2017, \mn@doi [\aj] {10.3847/1538-3881/aa6572}, \href
  {http://adsabs.harvard.edu/abs/2017AJ....153..215P} {153, 215}

\bibitem[\protect\citeauthoryear{{Petigura} et~al.,}{{Petigura}
  et~al.}{2017}]{2017AJ....153..142P}
{Petigura} E.~A.,  et~al., 2017, \mn@doi [\aj] {10.3847/1538-3881/aa5ea5},
  \href {http://adsabs.harvard.edu/abs/2017AJ....153..142P} {153, 142}

\bibitem[\protect\citeauthoryear{{Pollacco} et~al.,}{{Pollacco}
  et~al.}{2006}]{2006PASP..118.1407P}
{Pollacco} D.~L.,  et~al., 2006, \mn@doi [\pasp] {10.1086/508556}, \href
  {http://adsabs.harvard.edu/abs/2006PASP..118.1407P} {118, 1407}

\bibitem[\protect\citeauthoryear{{Ricker} et~al.,}{{Ricker}
  et~al.}{2016}]{2016SPIE.9904E..2BR}
{Ricker} G.~R.,  et~al., 2016, in Space Telescopes and Instrumentation 2016:
  Optical, Infrared, and Millimeter Wave. p. 99042B,
  \mn@doi{10.1117/12.2232071}

\bibitem[\protect\citeauthoryear{{Rickman} et~al.,}{{Rickman}
  et~al.}{2019}]{2019arXiv190401573R}
{Rickman} E.~L.,  et~al., 2019, arXiv e-prints, \href
  {http://adsabs.harvard.edu/abs/2019arXiv190401573R} {}

\bibitem[\protect\citeauthoryear{{Sestovic}, {Demory}  \& {Queloz}}{{Sestovic}
  et~al.}{2018}]{2018A&A...616A..76S}
{Sestovic} M.,  {Demory} B.-O.,   {Queloz} D.,  2018, \mn@doi [\aap]
  {10.1051/0004-6361/201731454}, \href
  {http://adsabs.harvard.edu/abs/2018A%26A...616A..76S} {616, A76}

\bibitem[\protect\citeauthoryear{{Spake} et~al.,}{{Spake}
  et~al.}{2018}]{2018Natur.557...68S}
{Spake} J.~J.,  et~al., 2018, \mn@doi [\nat] {10.1038/s41586-018-0067-5}, \href
  {http://adsabs.harvard.edu/abs/2018Natur.557...68S} {557, 68}

\bibitem[\protect\citeauthoryear{{Stassun} \& {Torres}}{{Stassun} \&
  {Torres}}{2018}]{2018ApJ...862...61S}
{Stassun} K.~G.,  {Torres} G.,  2018, \mn@doi [\apj]
  {10.3847/1538-4357/aacafc}, \href
  {http://adsabs.harvard.edu/abs/2018ApJ...862...61S} {862, 61}

\bibitem[\protect\citeauthoryear{{Szab{\'o}} \& {K{\'a}lm{\'a}n}}{{Szab{\'o}}
  \& {K{\'a}lm{\'a}n}}{2019}]{2019arXiv190304817S}
{Szab{\'o}} G.~M.,  {K{\'a}lm{\'a}n} S.,  2019, arXiv e-prints, \href
  {http://adsabs.harvard.edu/abs/2019arXiv190304817S} {}

\bibitem[\protect\citeauthoryear{{Temple} et~al.,}{{Temple}
  et~al.}{2018}]{2018MNRAS.480.5307T}
{Temple} L.~Y.,  et~al., 2018, \mn@doi [\mnras] {10.1093/mnras/sty2197}, \href
  {http://adsabs.harvard.edu/abs/2018MNRAS.480.5307T} {480, 5307}

\bibitem[\protect\citeauthoryear{{Thorngren} \& {Fortney}}{{Thorngren} \&
  {Fortney}}{2018}]{2018AJ....155..214T}
{Thorngren} D.~P.,  {Fortney} J.~J.,  2018, \mn@doi [\aj]
  {10.3847/1538-3881/aaba13}, \href
  {http://adsabs.harvard.edu/abs/2018AJ....155..214T} {155, 214}

\bibitem[\protect\citeauthoryear{{Triaud} et~al.,}{{Triaud}
  et~al.}{2013}]{2013A&A...551A..80T}
{Triaud} A.~H.~M.~J.,  et~al., 2013, \mn@doi [\aap]
  {10.1051/0004-6361/201220900}, \href
  {http://adsabs.harvard.edu/abs/2013A%26A...551A..80T} {551, A80}

\bibitem[\protect\citeauthoryear{{Udry} et~al.,}{{Udry}
  et~al.}{2019}]{2019A&A...622A..37U}
{Udry} S.,  et~al., 2019, \mn@doi [\aap] {10.1051/0004-6361/201731173}, \href
  {http://adsabs.harvard.edu/abs/2019A%26A...622A..37U} {622, A37}

\bibitem[\protect\citeauthoryear{{Van Eylen} et~al.,}{{Van Eylen}
  et~al.}{2016}]{2016AJ....152..143V}
{Van Eylen} V.,  et~al., 2016, \mn@doi [\aj] {10.3847/0004-6256/152/5/143},
  \href {http://adsabs.harvard.edu/abs/2016AJ....152..143V} {152, 143}

\bibitem[\protect\citeauthoryear{{Winn}}{{Winn}}{2010}]{2010arXiv1001.2010W}
{Winn} J.~N.,  2010, preprint, \href
  {http://adsabs.harvard.edu/abs/2010arXiv1001.2010W} {} (\mn@eprint {arXiv}
  {1001.2010})

\bibitem[\protect\citeauthoryear{{Winn} et~al.,}{{Winn}
  et~al.}{2010}]{2010ApJ...723L.223W}
{Winn} J.~N.,  et~al., 2010, \mn@doi [\apjl] {10.1088/2041-8205/723/2/L223},
  \href {http://adsabs.harvard.edu/abs/2010ApJ...723L.223W} {723, L223}

\makeatother
\end{thebibliography}


\appendix

\begin{table}
\renewcommand\thetable{A1}
\caption{WASP-166 Radial velocities.\protect\rule[-1.5mm]{0mm}{2mm}}
\begin{tabular}{cccr}
\hline
BJD\,--\,2400\,000 & RV & $\sigma_{\rm RV}$ & Bisector \\
(UTC)  & (km s$^{-1}$) & (km s$^{-1}$) & (km s$^{-1}$)\\ [0.5mm] \hline
\multicolumn{4}{l}{{\bf CORALIE:}}\\
56715.69870 & 	23.60674  &	0.00420  & $-$0.00118	\\
56718.61401 & 	23.60476  &	0.00449  & 0.00282	\\  [0.5mm] \hline
57186.46596 & 	23.59798  &	0.00571  & 0.00657	\\
57189.46427 & 	23.61726  &	0.00466  & 0.00915	\\
57379.74611 & 	23.60818  &	0.00352  & $-$0.00198\\
57420.73070 & 	23.58513  &	0.00363  & $-$0.00060\\
57421.63091 & 	23.59524  &	0.00374  & 0.00681	\\
57422.66015 & 	23.60186  &	0.00347  & $-$0.00721\\
57423.69579 & 	23.60940  &	0.00350  & 0.00762	\\
57426.68531 & 	23.58764  &	0.00402  & 0.01747	\\
57427.68034 & 	23.60654  &	0.00423  & $-$0.01649\\
57430.58045 & 	23.61098  &	0.00540  & 0.01034	\\
57431.68932 & 	23.59771  &	0.00449  & $-$0.01920\\
57432.65318 & 	23.59465  &	0.00399  & 0.00542	\\
57433.64406 & 	23.60229  &	0.00431  & 0.00174	\\
57434.76630 & 	23.60306  &	0.00587  & 0.01239	\\
57435.56404 & 	23.59491  &	0.00532  & 0.00404	\\
57451.62907 & 	23.61025  &	0.00390  & 0.00528	\\
57452.68003 & 	23.60150  &	0.00425  & 0.00954	\\
57453.59001 & 	23.60351  &	0.00427  & $-$0.00305\\
57458.58216 & 	23.59830  &	0.00395  & 0.00568	\\
57462.58709 & 	23.60401  &	0.00398  & 0.00156	\\
57488.57829 & 	23.61223  &	0.00397  & 0.00506	\\
57512.56673 & 	23.62081  &	0.00901  & 0.02664	\\
57558.45880 & 	23.59931  &	0.00409  & 0.00181	\\
57559.47429 & 	23.61804  &	0.00467  & 0.01462	\\
57560.45947 & 	23.60978  &	0.00461  & $-$0.00190\\
57561.46060 & 	23.60891  &	0.00490  & 0.00154	\\
57566.45959 & 	23.61192  &	0.00541  & 0.01347	\\
57568.46079 & 	23.59210  &	0.00522  & $-$0.00197\\
57716.80525 & 	23.62126  &	0.00707  & $-$0.00642\\
57721.78118 & 	23.60734  &	0.00450  & $-$0.01000\\
57722.79431 & 	23.61259  &	0.00426  & $-$0.00935\\
57723.79431 & 	23.61560  &	0.00480  & 0.00999	\\
57724.76502 & 	23.60758  &	0.00470  & $-$0.00170\\
57727.82177 & 	23.60882  &	0.00479  & 0.00605	\\
57730.78526 & 	23.59711  &	0.00498  & 0.00324	\\
57749.79985 & 	23.61638  &	0.00480  & 0.00369	\\
57758.72952 & 	23.59302  &	0.00425  & 0.00237	\\
57759.71655 & 	23.61154  &	0.00464  & 0.01173	\\
57762.68238 & 	23.60114  &	0.00523  & $-$0.00550\\  [0.5mm] \hline
\multicolumn{4}{l}{{\bf HARPS:}}\\
57486.54275 & 	23.61872  &	0.00217  & 0.01679 \\ 
57487.53361 & 	23.62463  &	0.00197  & 0.02873 \\ 
57523.49754 & 	23.62104  &	0.00324  & 0.02185 \\ 
57526.48623 & 	23.63539  &	0.00168  & 0.02013 \\ 
57536.54610 & 	23.63749  &	0.00174  & 0.03079 \\ 
57577.46098 & 	23.63466  &	0.00242  & 0.02971 \\ 
57728.82037 & 	23.63863  &	0.00157  & 0.02209 \\ 
57729.78720 & 	23.62939  &	0.00181  & 0.02392 \\ 
57730.82154 & 	23.62326  &	0.00187  & 0.02274 \\ 
57732.85980 & 	23.63434  &	0.00402  & 0.03581 \\ 
57733.78026 & 	23.63792  &	0.00185  & 0.02763 \\ 
57734.80841 & 	23.63362  &	0.00142  & 0.02294 \\ 
57735.83008 & 	23.62436  &	0.00144  & 0.02777 \\ 
57751.77429 & 	23.62868  &	0.00152  & 0.02248 \\ 
57752.73009 & 	23.62116  &	0.00160  & 0.02288 \\ 
57753.78115 & 	23.63266  &	0.00194  & 0.02552 \\ 
57754.80284 & 	23.64190  &	0.00182  & 0.01799 \\ 
57755.78101 & 	23.63761  &	0.00173  & 0.03517 \\ 
57756.85892 & 	23.62748  &	0.00163  & 0.02527 \\ [0.5mm] \hline
\end{tabular}
\end{table}

\begin{table}
\begin{tabular}{cccr}
\hline
BJD\,--\,2400\,000 & RV & $\sigma_{\rm RV}$ & Bisector \\
(UTC)  & (km s$^{-1}$) & (km s$^{-1}$) & (km s$^{-1}$)\\ [0.5mm] \hline
57767.58760 & 	23.63394  &	0.00303  & 0.01446	\\
57767.59207 & 	23.62533  &	0.00387  & 0.00627	\\
57767.60060 & 	23.62550  &	0.00393  & 0.02186	\\
57767.60469 & 	23.62297  &	0.00386  & 0.00775	\\
57767.60832 & 	23.61950  &	0.00287  & 0.02241	\\
57767.61151 & 	23.60388  &	0.00724  & 0.00137	\\
57767.61645 & 	23.63063  &	0.00339  & 0.02259 \\ 
57767.61977 & 	23.62539  &	0.00270  & 0.02510 \\ 
57767.62353 & 	23.62773  &	0.00259  & 0.02354 \\ 
57767.62751 & 	23.62473  &	0.00255  & 0.01744 \\ 
57767.63122 & 	23.62443  &	0.00249  & 0.02932 \\ 
57767.63503 & 	23.62001  &	0.00303  & 0.03186 \\ 
57767.63901 & 	23.62367  &	0.00298  & 0.01888 \\ 
57767.64282 & 	23.62850  &	0.00266  & 0.00989 \\ 
57767.64666 & 	23.62716  &	0.00240  & 0.02489 \\ 
57767.65047 & 	23.62837  &	0.00244  & 0.02723 \\ 
57767.65421 & 	23.62679  &	0.00287  & 0.01658 \\ 
57767.65819 & 	23.62576  &	0.00267  & 0.01895 \\ 
57767.66193 & 	23.62528  &	0.00299  & 0.02379 \\ 
57767.66586 & 	23.62671  &	0.00315  & 0.02334 \\ 
57767.66970 & 	23.62449  &	0.00263  & 0.02682	\\
57767.67342 & 	23.63029  &	0.00278  & 0.02365	\\
57767.67732 & 	23.62687  &	0.00279  & 0.02616	\\
57767.68104 & 	23.62274  &	0.00312  & 0.02647	\\
57767.68523 & 	23.62805  &	0.00326  & 0.02931	\\
57767.68907 & 	23.63029  &	0.00359  & 0.03063	\\
57767.69275 & 	23.62503  &	0.00343  & 0.01836	\\
57767.69634 & 	23.63200  &	0.00388  & 0.01038	\\
57767.70049 & 	23.63575  &	0.00401  & 0.02519	\\
57767.70436 & 	23.62679  &	0.00364  & 0.01344	\\
57767.70827 & 	23.63266  &	0.00397  & 0.01581	\\
57767.71208 & 	23.63109  &	0.00336  & 0.01289	\\
57767.71589 & 	23.62978  &	0.00322  & 0.02351	\\
57767.71979 & 	23.63458  &	0.00335  & 0.01285	\\
57767.72350 & 	23.63067  &	0.00344  & 0.00116	\\
57767.72728 & 	23.62371  &	0.00414  & 0.01412	\\
57767.73115 & 	23.63412  &	0.00569  & 0.01239	\\
57767.73531 & 	23.62115  &	0.00512  & 0.02570	\\
57767.73890 & 	23.62702  &	0.00385  & 0.00477	\\
57767.74261 & 	23.62558  &	0.00432  & 0.03501	\\
57767.74670 & 	23.62226  &	0.00344  & 0.01318	\\
57767.75044 & 	23.62675  &	0.00332  & 0.01451	\\
57767.75431 & 	23.61709  &	0.00378  & 0.03052	\\
57767.75819 & 	23.62384  &	0.00345  & 0.03771	\\
57767.76202 & 	23.62485  &	0.00329  & 0.01941	\\
57767.76605 & 	23.62461  &	0.00340  & 0.02824	\\
57767.76967 & 	23.61390  &	0.00337  & 0.02390	\\
57767.77348 & 	23.61676  &	0.00333  & 0.02542	\\
57767.77738 & 	23.61591  &	0.00350  & 0.02449	\\
57767.78119 & 	23.61169  &	0.00320  & 0.02376	\\
57767.78500 & 	23.60689  &	0.00326  & 0.02685	\\
57767.78895 & 	23.61830  &	0.00339  & 0.02571	\\
57767.79275 & 	23.61151  &	0.00309  & 0.02777	\\
57767.79656 & 	23.61055  &	0.00319  & 0.01576	\\
57767.80037 & 	23.61029  &	0.00323  & 0.01117	\\
57767.80428 & 	23.61641  &	0.00314  & 0.01905	\\
57767.80822 & 	23.60975  &	0.00364  & 0.03807	\\
57767.81189 & 	23.61252  &	0.00336  & 0.02892	\\
57767.81577 & 	23.61281  &	0.00348  & 0.00388	\\
57767.81957 & 	23.61239  &	0.00349  & 0.01962	\\
57767.82342 & 	23.62330  &	0.00302  & 0.01296	\\
57767.82726 & 	23.62029  &	0.00311  & 0.01471	\\
57767.83121 & 	23.61806  &	0.00296  & 0.03654	\\
57767.83504 & 	23.62090  &	0.00278  & 0.01999	\\
57767.83881 & 	23.62567  &	0.00272  & 0.02272	\\
\end{tabular}
\end{table}

\begin{table}
\begin{tabular}{cccr}
\hline
BJD\,--\,2400\,000 & RV & $\sigma_{\rm RV}$ & Bisector \\
(UTC)  & (km s$^{-1}$) & (km s$^{-1}$) & (km s$^{-1}$)\\ [0.5mm] \hline
57767.84263 & 	23.62769  &	0.00276  & 0.03642	\\
57767.84646 & 	23.62409  &	0.00290  & 0.02824	\\
57767.85041 & 	23.62141  &	0.00297  & 0.02191	\\
57767.85429 & 	23.62807  &	0.00285  & 0.02369	\\
57767.85802 & 	23.61976  &	0.00308  & 0.02249 \\ 
57767.86187 & 	23.62543  &	0.00292  & 0.03571 \\ 
57767.86566 & 	23.61653  &	0.00328  & 0.01006 \\ 
57767.86958 & 	23.62595  &	0.00320  & 0.03263 \\ 
57767.87338 & 	23.62471  &	0.00306  & 0.02682 \\ 
57767.87724 & 	23.62356  &	0.00284  & 0.02180 \\ [0.5mm] \hline
57769.80714 & 	23.61499  &	0.00219  & 0.02590 \\ 
57770.75537 & 	23.62649  &	0.00281  & 0.02549 \\ [0.5mm] \hline
57816.65537 & 	23.63019  &	0.00210  & 0.02144	\\
57816.66032 & 	23.62917  &	0.00204  & 0.01473	\\
57816.67017 & 	23.62837  &	0.00244  & 0.01715	\\
57816.67507 & 	23.63615  &	0.00262  & 0.01374	\\
57816.67886 & 	23.63392  &	0.00261  & 0.02071	\\
57816.68266 & 	23.63469  &	0.00283  & 0.01612	\\
57816.68652 & 	23.63993  &	0.00280  & 0.01991	\\
57816.69042 & 	23.63850  &	0.00285  & 0.02664	\\
57816.69419 & 	23.63915  &	0.00271  & 0.02256	\\
57816.69791 & 	23.63940  &	0.00279  & 0.01625	\\
57816.70188 & 	23.63349  &	0.00323  & 0.03297	\\
57816.70578 & 	23.63760  &	0.00310  & 0.02489	\\
57816.70961 & 	23.63197  &	0.00300  & 0.01579	\\
57816.71334 & 	23.63354  &	0.00289  & 0.01285	\\
57816.71721 & 	23.63807  &	0.00281  & 0.02084	\\
57816.72097 & 	23.63601  &	0.00292  & 0.02362	\\
57816.72490 & 	23.63239  &	0.00297  & 0.01569	\\
57816.72856 & 	23.63962  &	0.00312  & 0.01805	\\
57816.73260 & 	23.63272  &	0.00334  & 0.01253	\\
57816.73633 & 	23.62798  &	0.00317  & 0.02418	\\
57816.74019 & 	23.62777  &	0.00325  & 0.02878	\\
57816.74404 & 	23.62656  &	0.00312  & 0.02588	\\
57816.74787 & 	23.62564  &	0.00335  & 0.03502	\\
57816.75177 & 	23.62932  &	0.00324  & $-$0.00833\\
57816.75556 & 	23.62015  &	0.00320  & 0.03667	\\
57816.75929 & 	23.62272  &	0.00305  & 0.03631	\\
57816.76336 & 	23.62788  &	0.00344  & 0.02356	\\
57816.76706 & 	23.62189  &	0.00314  & 0.01721	\\
57816.77082 & 	23.61831  &	0.00307  & 0.04728	\\
57816.77461 & 	23.62033  &	0.00303  & 0.01428	\\
57816.77864 & 	23.61918  &	0.00347  & 0.03644	\\
57816.78237 & 	23.61925  &	0.00330  & 0.03297	\\
57816.78610 & 	23.61773  &	0.00355  & 0.02491	\\
57816.79007 & 	23.62124  &	0.00328  & 0.02607	\\
57816.79383 & 	23.61770  &	0.00337  & 0.02593	\\
57816.79766 & 	23.61668  &	0.00384  & 0.03799	\\
57816.80145 & 	23.60814  &	0.00352  & 0.02619	\\
57816.80537 & 	23.61670  &	0.00362  & 0.02351	\\
57816.80909 & 	23.61294  &	0.00376  & 0.02974	\\
57816.81303 & 	23.62676  &	0.00385  & 0.01295	\\
57816.81682 & 	23.62619  &	0.00381  & 0.01807	\\
57816.82062 & 	23.61808  &	0.00393  & 0.02724 \\ 
57816.82449 & 	23.62186  &	0.00398  & 0.03071 \\ 
57816.82832 & 	23.62480  &	0.00416  & 0.02181 \\ 
57816.83212 & 	23.62562  &	0.00439  & 0.02305 \\ 
57816.83599 & 	23.61783  &	0.00442  & 0.00742 \\ 
57816.83982 & 	23.62403  &	0.00471  & 0.02442 \\ 
57816.84358 & 	23.62622  &	0.00465  & 0.01361 \\ 
57816.84738 & 	23.61928  &	0.00512  & $-$0.00844	\\
57816.85131 & 	23.61671  &	0.00493  & 0.02648 \\ 
57816.85511 & 	23.62293  &	0.00502  & 0.02198 \\ 
57816.85898 & 	23.61311  &	0.00498  & 0.00126 \\ [0.5mm] \hline
\end{tabular}
\end{table}

\begin{table}
\begin{tabular}{cccr}
\hline
BJD\,--\,2400\,000 & RV & $\sigma_{\rm RV}$ & Bisector \\
(UTC)  & (km s$^{-1}$) & (km s$^{-1}$) & (km s$^{-1}$)\\ [0.5mm] \hline
57827.53425 & 	23.62930  &	0.00280  & 0.02204	\\
57827.53872 & 	23.63603  &	0.00290  & 0.02670	\\
57827.54345 & 	23.63423  &	0.00285  & 0.02590	\\
57827.54786 & 	23.62957  &	0.00259  & 0.02816	\\
57827.55227 & 	23.63447  &	0.00278  & 0.01389	\\
57827.55676 & 	23.63066  &	0.00297  & 0.02514	\\
57827.56098 & 	23.63909  &	0.00305  & 0.02265	\\
57827.56519 & 	23.63729  &	0.00373  & 0.02886	\\
57827.57028 & 	23.64419  &	0.00428  & 0.02696	\\
57827.57457 & 	23.64038  &	0.00364  & 0.02522	\\
57827.57890 & 	23.65067  &	0.00382  & 0.02796	\\
57827.58327 & 	23.64449  &	0.00309  & 0.02050	\\
57827.58769 & 	23.64204  &	0.00295  & 0.02194	\\
57827.59202 & 	23.64369  &	0.00286  & 0.03653	\\
57827.59643 & 	23.64428  &	0.00285  & 0.02834	\\
57827.60084 & 	23.64191  &	0.00267  & 0.03241	\\
57827.60533 & 	23.63830  &	0.00263  & 0.01467	\\
57827.60976 & 	23.64247  &	0.00262  & 0.02718	\\
57827.61409 & 	23.64307  &	0.00264  & 0.01784	\\
57827.61850 & 	23.63863  &	0.00266  & 0.04245	\\
57827.62295 & 	23.63842  &	0.00252  & 0.01951	\\
57827.62728 & 	23.63858  &	0.00241  & 0.02096	\\
57827.63170 & 	23.63444  &	0.00238  & 0.02134	\\
57827.63615 & 	23.63524  &	0.00238  & 0.01689	\\
57827.64056 & 	23.63139  &	0.00264  & 0.01744	\\
57827.64497 & 	23.63212  &	0.00243  & 0.02347	\\
57827.64938 & 	23.62593  &	0.00239  & 0.02557	\\
57827.65372 & 	23.62390  &	0.00251  & 0.02587	\\
57827.65821 & 	23.62888  &	0.00241  & 0.02687	\\
57827.66262 & 	23.62513  &	0.00246  & 0.02387	\\
57827.66703 & 	23.61826  &	0.00244  & 0.03302	\\
57827.67144 & 	23.62351  &	0.00237  & 0.01251	\\
57827.67526 & 	23.60691  &	0.00768  & $-$0.01366\\
57827.68085 & 	23.61397  &	0.00330  & 0.02234	\\
57827.68469 & 	23.62346  &	0.00271  & 0.01181	\\
57827.68917 & 	23.61960  &	0.00271  & 0.01719	\\
57827.69354 & 	23.62385  &	0.00272  & 0.01891	\\
57827.69801 & 	23.62317  &	0.00337  & 0.03128	\\
57827.70243 & 	23.62472  &	0.00259  & 0.02089	\\
57827.70685 & 	23.62682  &	0.00261  & 0.01774	\\
57827.71126 & 	23.63026  &	0.00259  & 0.02072	\\
57827.71559 & 	23.62902  &	0.00272  & 0.00473	\\
57827.71986 & 	23.62284  &	0.00418  & 0.02883	\\
57827.72448 & 	23.63551  &	0.00267  & 0.03343	\\
57827.72884 & 	23.62753  &	0.00266  & 0.02473	\\
57827.73326 & 	23.63365  &	0.00270  & 0.02559	\\
57827.73771 & 	23.62836  &	0.00282  & 0.02362	\\
57827.74242 & 	23.62463  &	0.00315  & 0.02299	\\
57827.74657 & 	23.63381  &	0.00288  & 0.02348	\\
57827.75090 & 	23.62604  &	0.00292  & 0.01722	\\
57827.75536 & 	23.62729  &	0.00298  & 0.03213	\\
57827.75977 & 	23.62853  &	0.00294  & 0.02042	\\
57827.76467 & 	23.63083  &	0.00395  & 0.02306	\\
57827.76868 & 	23.63494  &	0.00308  & 0.04233	\\
57827.77297 & 	23.61858  &	0.00347  & 0.02458	\\
57827.77742 & 	23.63191  &	0.00324  & 0.01560	\\
57827.78183 & 	23.63586  &	0.00356  & 0.01705 \\ 
57827.78629 & 	23.62552  &	0.00359  & 0.02208 \\ 
57827.79066 & 	23.63287  &	0.00373  & 0.01926 \\ 
57827.79507 & 	23.62353  &	0.00402  & 0.02713 \\ 
57827.79944 & 	23.63074  &	0.00404  & 0.02189 \\ 
57827.80389 & 	23.63272  &	0.00406  & 0.01771 \\ 
57827.80837 & 	23.62434  &	0.00435  & $-$0.00767	\\
57827.81284 & 	23.62679  &	0.00437  & 0.02713 \\ 
57827.81780 & 	23.62169  &	0.00599  & 0.03090 \\ 
57827.82180 & 	23.63132  &	0.00444  & 0.01595 \\ [0.5mm] \hline
\end{tabular}
\end{table}

\begin{table}
\begin{tabular}{cccr}
\hline
BJD\,--\,2400\,000 & RV & $\sigma_{\rm RV}$ & Bisector \\
(UTC)  & (km s$^{-1}$) & (km s$^{-1}$) & (km s$^{-1}$)\\ [0.5mm] \hline
58198.71811 & 	23.61212  &	0.00130  & 0.02011 \\ 
58199.67572 & 	23.61200  &	0.00111  & 0.02893 \\ 
58201.64627 & 	23.63373  &	0.00139  & 0.02105 \\ 
58202.70438 & 	23.63717  &	0.00119  & 0.01943 \\ 
58203.61226 & 	23.62522  &	0.00141  & 0.01834 \\ 
58204.60885 & 	23.61929  &	0.00145  & 0.02213 \\ [0.5mm] \hline
\multicolumn{4}{l}{Bisector errors are twice RV errors. Lines denote an}\\
\multicolumn{4}{l}{instrument upgrade, or separate through-transit sequences}
\\
\end{tabular}
\end{table}

\bsp	
\label{lastpage}
\end{document}